\documentclass[twocolumn,letterpaper,amsmath,amssymb,floatfix,aps,superscriptaddress]{revtex4-1}

\usepackage{hyperref}
\usepackage{verbatim}
\usepackage{graphicx}
\usepackage{bm} 
\usepackage{ifpdf} 
\usepackage{dcolumn}  
\usepackage{epsfig}
\usepackage{epstopdf} 
\usepackage{color}
\usepackage[usenames,dvipsnames]{xcolor}

\begin{document}

\title{Transport coefficients for hard-sphere relativistic gas}

\author{Malihe Ghodrat}
\thanks{Email: \texttt{m.ghodrat@modares.ac.ir}}
\affiliation{Department of Physics, Faculty of Basic Sciences, Tarbiat Modares University, P.O. Box 14115-175, Tehran, Iran}



\begin{abstract}
Transport coefficients are of crucial importance in theoretical as well as experimental studies. Despite substantial research on classical hard sphere/disk gases in low and high density regimes, a thorough investigation of transport coefficients for massive relativistic systems is missing in the literature. In this work a fully relativistic molecular dynamics simulation is employed to numerically obtain the transport coefficients of a hard sphere relativistic gas based on Helfand-Einstein expressions. The numerical data are then used to check the accuracy of Chapmann-Enskog (CE) predictions in a wide range of temperature. The results indicate that while simulation data in low temperature regime agrees very well with theoretical predictions, it begins to show deviations as temperature rises, except for the thermal conductivity which fits very well to CE theory in the whole range of temperature. Since our simulations are done in low density regimes, where CE approximation is expected to be valid, the observed deviations can be attributed to the inaccuracy of linear CE theory in extremely relativistic cases.
\end{abstract}


\maketitle
\section{Introduction}
Transport coefficients characterize the dissipative mechanisms which bring a disturbed system to its equilibrium state. In macroscopic description these coefficients appear in phenomenological transport equations for transfer of mass (Fick's law of diffusion), energy (Fourier's law of heat conduction) and momentum (Newton's law of viscosity) as well as in Navier-Stokes equations, to relate thermodynamic forces (density, energy, momentum or pressure gradients) to their corresponding fluxes.

On a more fundamental level, the hydrodynamic equations of a fluid and explicit expressions for transport coefficients are obtained by solving Boltzmann transport equation with an appropriate expansion of density distribution function around a local equilibrium state. A systematic method in this direction was first introduced independently by Chapman and Enskog in which the ratio of mean free path to a typical macroscopic length is used as the expansion parameter. The subsequent degrees of approximation would then lead to Euler equations, Navier-Stokes equations, Burnett equations and so on \cite{Rei80}.

Along with advances in description of classical systems, Boltzmann kinetic theory was extended to the domain of relativity in early 1900s \cite{Jut11,Lich40}. High temperature systems of fast moving particles (e.g. quark-gluon plasmas) as well as moderate-temperature flows moving at extremely high velocities (e.g supernova explosions, the cosmic expansion, and solar flares) are the two main classes where relativistic corrections become non-negligible. The adoption of CE approximation in this context \cite{Isr63,Kel63,Cher64}, made it possible to obtain the corresponding transport coefficients for relativistic fluid \cite{Gro80,Cer02}. Since then, the relativistic kinetic theory has been used in a wide range of phenomena from large scale scenarios studying the evolution of the universe and formation of galaxies \cite{Rez13,Lop11,Bal01} to subatomic scales in heavy-ion collision experiments at CERN and BNL \cite{Huo04,Kol04} to recent applications in graphene studies \cite{Mul08a,Mul09,Men11,Tor15}.     

A common feature shared by phenomenological macroscopic laws and the first CE approximation is that the relation between stimuli (or forces) and the resulting fluxes (responses) are linear, with the ``transport coefficients" as their proportionality constants. This assumption, which remains unchanged in the relativistic CE theory, is basically inconsistent with the fundamental  assumptions of special relativity; in the sense that it leads to first-order hydrodynamic theories in which the superluminal propagation of fluctuations are possible due to the parabolic nature of the dynamical equations \cite{Rez13}. The discrepancy still remains in higher order Burnett and superBurnet theories, which are mostly used in the linearized form; e.g. in theory of sound \cite{Gro80}. This fundamental problem has led to relativistic \cite {Isr76,Stew77,Liu86} (and even nonrelativistic \cite{Muel67,Liu83}) extended second order theories. 

In addition to extended theories, attempts have been made in the context of first-order theories  to solve the problems associated with relativistic fluids \cite{Gcol06,Gper09b,Gper10,Tsum07,Van12}. Recent numerical studies have also shown that linear transport laws and first order theories give a good description of equilibration processes \cite{Gho13} and propagation of fluctuations \cite{Gho16} in hydrodynamic limit. The question is to what extent one can rely on the transport coefficients obtained from linear CE method?    

To provide a possible answer, we refer to the fundamental development in the theory of transport processes which made it possible to calculate the transport coefficients in terms of the microscopic properties. The method, best known as Green-Kubo \cite{Gre51,Kub57}, derives expressions for transport coefficients based on equilibrium time-correlation functions. In the pioneering works by Alder and co-workers \cite{Ald70}, molecular dynamics simulations were conducted to calculate the viscosity of a hard sphere gas. Their algorithm, in fact, uses Helfand generalization \cite{Hel60} of Einstein relation for self-diffusion coefficient, the so called  Helfand-Einstein relations, which are derived from Green-Kubo formulas. Since then, the validity of classical CE description for hard sphere/disk gasseous systems and other fluids has been widely studied in low and moderate density regimes \cite{Gar06, Vis07}. 

The objective of this work is to study the accuracy of transport coefficient predicted by relativistic Chapman-Enskog approximation, comparing them to numerical transport coefficients based on Einstein-Helfand method. It specifically focuses on low density systems, where linear first order theories are expected to give reliable description of the fluid, so that the impact of relativistic effects in high temperature limit become detectable.

To this end, the manuscript is organized as follows. In section \ref{s:TB}, the theoretical background is given focusing on the derivation of transport coefficients based on two methods, first the CE approximation and second the time correlation functions and their equivalent Einstein-Helfand relations. Section \ref{s:MST} devotes to the description of our model system, computational methods to calculate transport coefficients, and how to take into account the limitations caused by periodic boundary conditions. In section \ref{s:R} the numerical results are compared to the CE predictions. We'll finally conclude in section \ref{s:C}.

\section{Theoretical background} \label {s:TB}
\subsection{Relativistic Chapman-Enskog method}
The generalization of statistical mechanics to the domain of relativity was first established by J\"{u}ttner in 1911 when he succeeded to derive the relativistic form of Maxwell-Boltzmann distribution \cite{Jut11}. With the development of relativistic kinetic (transport) equations and \textit{H}-theorem, the conservation laws of mass and energy-momentum as well as the law of entropy production were obtained. Finally, it was the adoption of Chapman-Enskog \cite{Chap16} and Maxwell-Grad \cite{Gra49} methods for solving the kinetic equation, that made it possible to determine expressions for transport coefficients of relativistic fluids.  

The relativistic CE method, which is our focus in this work, is a straightforward generalization of the classical version aiming to solve the transport equation in hydrodynamic limit, where the system is in a local equilibrium and the non-uniformities slowly relaxes to global equilibrium. In this method, the density distribution is expanded around the equilibrium distribution with the ratio of mean free path to a typical macroscopic length scale $\epsilon$ as the expansion parameter \cite{Rei80,Gro80}. 
\begin{equation}
f=f^{(0)}+\epsilon f^{(1)} +\epsilon^2 f^{(2)}+\dots
\end{equation}
CE method relies on the hydrodynamic limit where $\epsilon$ is small enough to make the first approximation accurate. The crucial assumption of the method is that the density distribution function, $f$, can be expressed as a function of only hydrodynamic variables and their gradients. So, 
\begin{equation}
f(x,p)=f^{(0)}(x,p)[1+\phi(x,p)]
\end{equation}
with equilibrium density function 
\begin{equation}
f^{(0)}(x,p)=\frac{1}{2\pi\hbar^3}\frac{1}{\exp[-\beta(\mu-p^\nu U_\nu (x))]\pm a},
\end{equation}
in which $e^{\beta\mu}$ is fugacity, and $a=0,-1,+1$ gives classical, Fermi-Dirac and Bose-Einstein statistics, respectively. 
The deviation function, $\phi(x,p)$, is given by
\begin{equation}
\phi(x,p)=\frac{1}{c n \sigma (T)}(A X -cB_\mu X_q^\mu + C^{\mu\nu} \mathring{X}_{\mu\nu}),
\end{equation}
in which $X=-\nabla^\mu U_\mu$, $X_q^\mu=\nabla^\mu \log T-\frac{k_B T}{h} \nabla^\mu \log P$ and $ \mathring{X}_{\mu\nu}=(\Delta_\sigma^\mu\Delta_\tau^\nu-\frac{1}{3}\Delta^{\mu\nu}\Delta_{\sigma\tau})\nabla^\sigma U^\tau$ are thermodynamics forces, with $h$ representing the enthalpy per particle. $A$, $B$, and $C$ are dimensionless quantities expressed in terms of momentum vector $p^\mu$, metric tensor $g^{\mu\nu}$, and thermodynamic variables: density $n$, temperature $T$ and hydrodynamic velocity $U^\mu$. We recall that $\Delta{\mu\nu}=g^{\mu\nu}-c^{-2}U^\mu U^\nu$ is the projection operator, $\sigma(T)$ is the characteristic cross section and  $c$ denotes the speed of light.

Once the CE first approximation is used in relativistic kinetic equation and the method is applied, the explicit expression for transport coefficients is given in terms of particle interactions, which would reduce to the following expressions (diffusion coefficient, shear and bulk viscosity, and thermal conductivity, respectively) for the case of hard-sphere gas \cite{Gro80}:
\begin{eqnarray}
D&=&\frac{3}{16\pi}\frac{c}{ n\sigma_{12}}\frac{z^2K_2^2(z)}{(2z^2+1)K_2(2z)+7zK_3(2z)}\label{e:CE_D}\\
\nonumber\eta_s&=&\frac{15}{32\pi}\frac{k_B T}{c\sigma}\frac{z^2K_2^2(z)\hat h^2}{(15z^2+2)K_2(2z)+(3z^2+49z)K_3(2z)} \\
&&\label{e:CE_etas}\\
\eta_b&=&\frac{1}{32\pi}\frac{k_B T}{c\sigma}\frac{z^2 K_2^2(z)[(5-3\gamma)\hat h-3\gamma]^2}{2K_2(2z)+zK_3(2z)} \label{e:CE_etab}\\
\lambda&=& \frac{3}{32\pi}\frac{c k_B}{\sigma}\frac{z^2 K_2^2(z)[\gamma/(\gamma-1)]^2}{(z^2+2)K_2(2z)+5zK_3(2z)}
\label{eq:CE_lambda}\end{eqnarray}
with $z=mc^2/k_B T$, $\hat h= z h=z K_3(z)/K_2(z)$, $
\gamma/(\gamma-1)= z^2+5\hat h -\hat h^2$. $K_n$ is the modified Bessel function of order $n$, $\sigma=2R^2$ and $\sigma_{12}=R^2$ are the relevant cross sections, $R$ and $m$ referring to the radius and mass of the particles. Transport coefficients can also be derived based on relaxation time models of Boltzmann equation. Description of the model and a comparison to the above expressions are given in \cite{Cer01,And74}.

\subsection{Time correlation functions and transport coefficients}
The dissipative phenomena that occur out of equilibrium are the same mechanisms that govern the decay of fluctuations about the equilibrium state (fluctuation-dissipation theorem \cite{Cal51}). This makes it possible to measure the transport coefficients either by conducting a suitable non-equilibrium experiment \cite{Ash73} or making observation of the fluctuating quantities associated with each transport coefficient using Green-Kubo or Einstein-Helfand relations \cite{Hel60}.

The so called Green-Kubo formulas give each transport coefficient, $\alpha$, as the time integral of the autocorrelation of a specific microscopic flux, $\mathcal {\dot A}$,
\begin{equation}
\alpha=\int_0^\infty dt \left\langle \mathcal {\dot A }(t)\mathcal{\dot A}(0)\right\rangle,
\end{equation}\label{eq:Corr}
while their equivalent `Einstein-Helfand' relations 
\begin{equation}
2t\alpha=\left\langle |\mathcal {A}(t)-\mathcal {A}(0)|^2\right\rangle=\left\langle|\delta\mathcal {A}(t)|^2\right\rangle,
\end{equation}\label{eq:Ein-Hel} 
are based on Einstein classical work on Brownian motion, which relates the second moment of the displacement to diffusion coefficient, and Helfand generalization of Einstein formula for other coefficients.

Since Einstein-Helfand formulas do not involve the forces, they are considered more appropriate to be used in the case of hard sphere/disk interactions, where the instantaneous forces between particles are ill-defined and would lead to singular contributions especially at high densities \cite{All86}. However, the fluctuating quantity, $\mathcal A$ in Einstein-Helfand relation, is calculated on the real trajectory of a particle that is not accessible given the periodic boundary condition applied in MD simulations. To overcome this problem, we have adopted the method proposed in \cite{Gar06} where the increment in $\mathcal {A}$, is measured in well defined time intervals instead of calculating its value at successive times. In this case, $\delta \mathcal {A}$ might be divided into `\emph{kinetic}' and `\emph{collision}' parts
\begin{equation}
\delta\mathcal {A}=\delta\mathcal {A}^{(K)}+\delta\mathcal {A}^{(C)}.
\end{equation}
The idea is that in a time interval $[t,t+\delta t]$ where no collision occurs, the contribution in $\delta\mathcal {A}$ is purely due to the displacement of the particle, while the discontinuous change of the velocities in collisions gives an additional instantaneous jump in $\delta \mathcal {A}$, which has been taken into account in `collision' term. More detail on this method is given in section \ref{s:MST}, where we further elaborate on the calculation techniques of each transport coefficient.

\section{Model and Simulation Techniques}\label {s:MST}
Our model simulates a three dimensional hard-sphere relativistic gas using an event-driven molecular dynamics method. In this model, $N$ hard-sphere particles of radius $R$ and rest mass $m$ are enclosed in a box of volume $V$ with periodic boundary conditions. The inter-particle interactions are purely repulsive, and only occur at center-to-center distance of $2R$, otherwise the particles move in straight lines. The collisions are governed by relativistic energy-momentum conservation laws, assuming that momentum is only transferred in the direction of connecting line between the centers (i.e, elastic head-on collisions) \cite{Gho11,Gho13}. The interaction cross section associated with the hard-sphere model is independent of the energy and of the scattering angle. This makes it a suitable choice to simulate massive hadrons whose total cross sections are more or less constant in the energy range of interest in relativistic kinetic theory \cite{Gro80}. The temperature of the system is related to the total energy per particle by the expression \cite{Mon09}
\begin{equation}
e=\langle E\rangle/N=3 k_B T+\frac{K_1(z)}{K_2(z)},
\end{equation}
We have adopted natural units ($k_B=c=1$) and set $m=1$ in simulations, such that the inverse temperature parameter $z^{-1}=T$. To have an estimate of the actual temperatures, one might take hard spheres with neutron mass, which gives a temperature of $10^{13} K$ for $z^{-1}=1$. This means that even weakly relativistic regime (e.g. $z^{-1}=0.01$) corresponds to a fairly high temperature, a fact which puts serious limits on our access to experimental data in relativistic regime and justifies the necessity of simulations and/or numerical studies in this field. 

The model presented above, is a fully relativistic one without adjustable parameters or probabilistic factors that can be employed as a reliable numerical laboratory. So far, it has been successfully used to investigate thermostatistical \cite{Gho11,Gho13} and hydrodynamic \cite{Gho16} properties  of relativistic fluids in low density regime. In this work we tend to obtain the transport coefficients using Einstein-Helfand relations and compare simulation results to theoretical predictions based on Chapman-Enskog method. The systems under study are all in low density regime ($\rho\sim0.02$), where linear CE theory is believed to be a good approximation, while temperature parameter covers a wide range, from low temperature classical regime to extremely relativistic limit.  

\subsection{Self-diffusion coefficient}\label{ss:SD}
\begin{figure}
	\centering
	\includegraphics[width=0.75\linewidth]{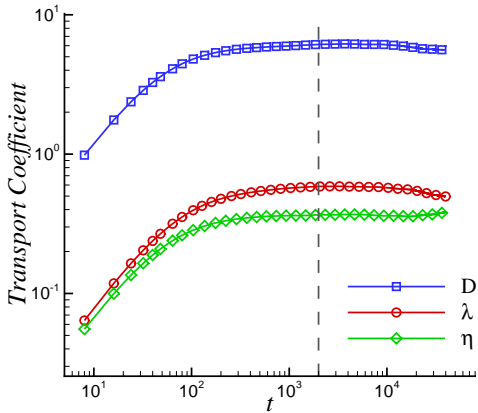}
	\caption{Diffusion coefficients ($D$), thermal conductivity ($\lambda$) and shear viscosity ($\eta_s$) are plotted versus time for a system of $N=1000$, $\rho=0.018$ and $T=1$, using Einstein-Helfand relations with time series of length $N_{run}=10^7$. After a transient regime in lower time scales, the coefficients saturate to a fixed value (highlighted by dashed line) which we report as ``numerical transport coefficients". The roughness in tails goes back to the insufficient averaging due to the finite length of time series.}
	\label{fig:Trans_t}
\end{figure}

Under a steady state condition the flux of mass is linearly related to the gradient of concentration with diffusion coefficient, $D$, as proportionality constant. 
\begin{equation}
\mathbf{J}_n=-D\mathbf{\nabla} n
\end{equation} 
Best known as Fick's law, this equation forms the core of our understanding of diffusion in solids, liquids, and gases. Equivalently, one might obtain the diffusion coefficient (in $d$ dimensions) based on `Einstein-Helfand' relation,
\begin{equation}
D=\frac{1}{2d}\lim_{t\rightarrow\infty}\frac{d}{dt}\left\langle | {\mathbf r}(t)- {\mathbf r}(0)|^2\right\rangle,
\label{eq:Diff_Coef}\end{equation}
in which $\mathbf r(t)$ is the position of tagged particle at time $t$, and $\left\langle ...\right\rangle$ denote ensemble averaging in equilibrium.

In order to numerically calculate Eq.\ref{eq:Diff_Coef}, we measure the displacements of a tagged particle from its original position at equal intervals of time, $\Delta t$, which is typically a small multiple of simulation time step (or average time step in case of event driven MD) and less than the mean free time of the system. This, will give us a time series of length $N_{run}$, that can be used to obtain the second moment of displacement averaged over $N_{max}$ time origins, $t_n$.
\begin{equation}
\left\langle |{\mathbf r}(t)-{\mathbf r}(0)|^2\right\rangle=\frac{1}{N N_{max}} \sum_{i=1}^{N_{max}}\sum_{n=1}^{N}|{\mathbf r_i}(t+t_n)-{\mathbf r_i}(t_n)|^2
\label{eq:Diff_Coef_Sim}\end{equation}

The value of $N_{max}$ depends on $t$. For $t=\Delta t$, one can average over  $N_{max}=N_{run}$ time origins while the statistics becomes poorer as $t$ increases, such that for longest time, $t=N_{run}\Delta t$, there would be only one term in the summation, i.e. $N_{max}=1$. 

This averaging method \cite{All86}, which will also be used in the following sections, gives the transport coefficients as a function of time, $t$. Here, the boundaries and periodicity of the system are not problematic issues, because we're keeping the sum of displacements which are measured at each time step. Additionally, diffusion coefficient is a single-particle property, and since all particles are equal, one can also average over all particles present in the system to gather larger statistics (See Eq.~\ref{eq:Diff_Coef_Sim}).    

\begin{figure*}[t!]
	\begin{center}
		\begin{minipage}[h]{0.32\textwidth}	\begin{center}
				\includegraphics[width=\textwidth]{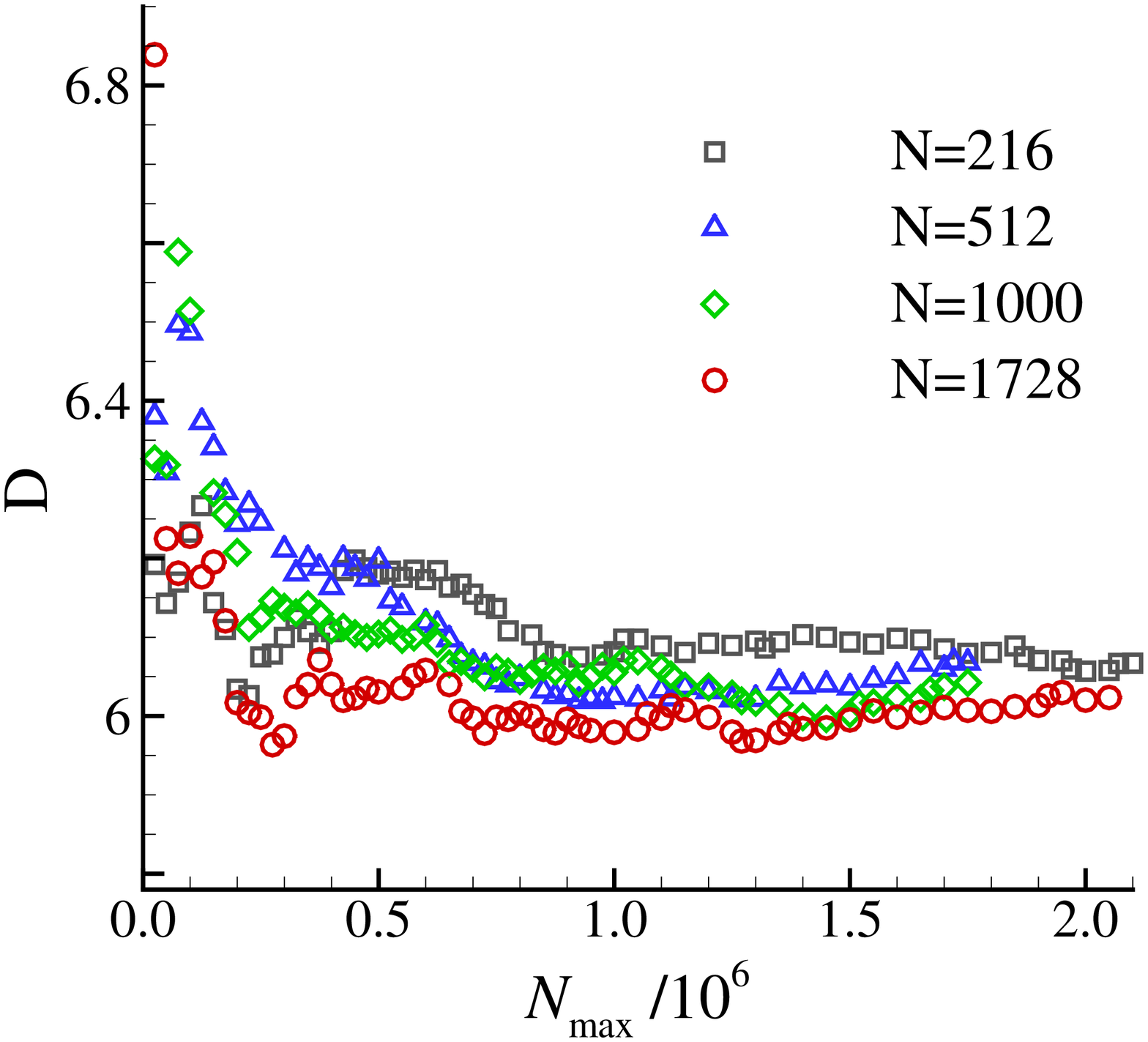} (a)
		\end{center} \end{minipage} \hskip0.01cm	
		\begin{minipage}[h]{0.32\textwidth}	\begin{center}
				\includegraphics[width=\textwidth]{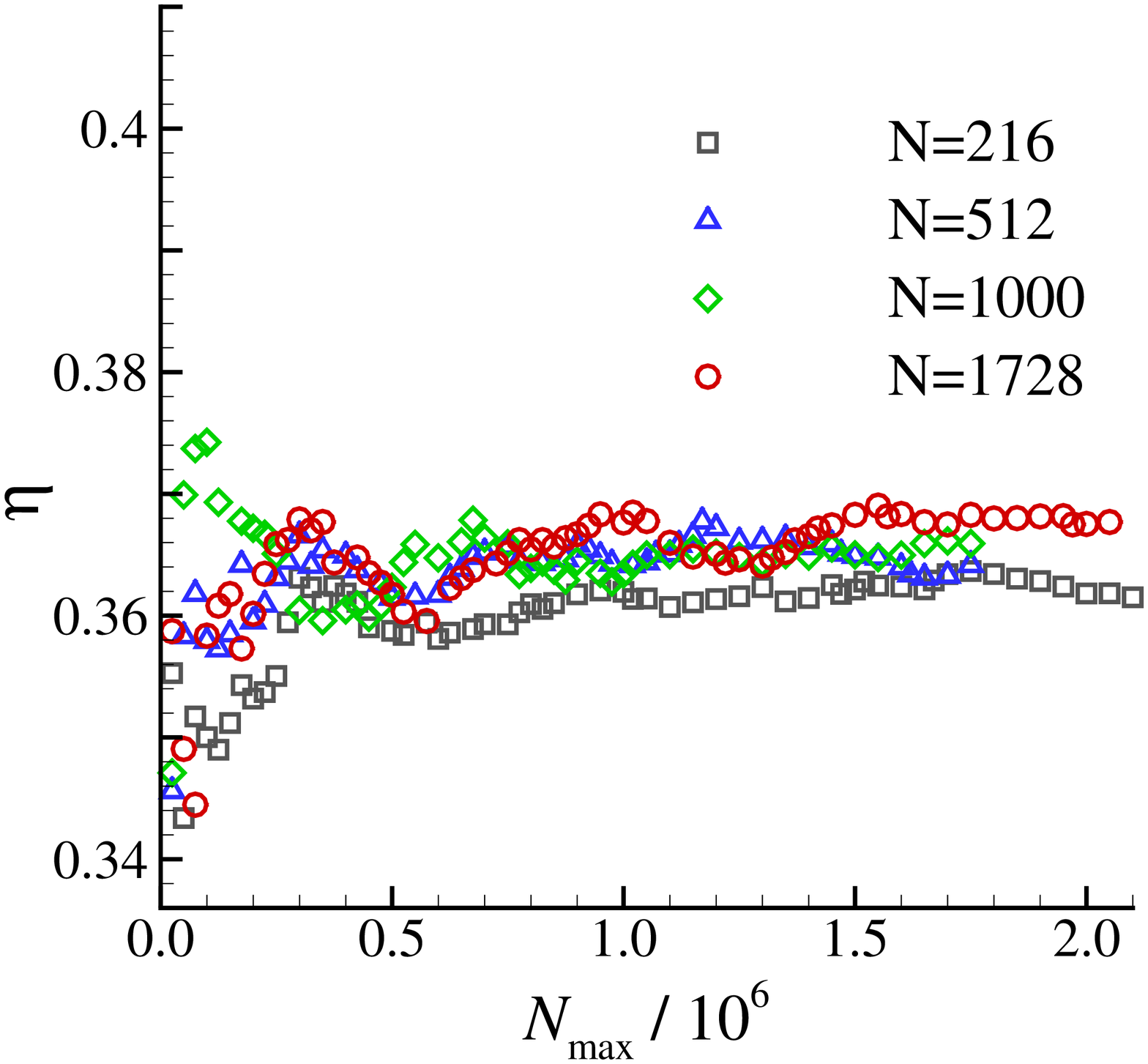} (b)
		\end{center} \end{minipage} \hskip0.01cm	
		\begin{minipage}[h]{0.32\textwidth}\begin{center}
				\includegraphics[width=\textwidth]{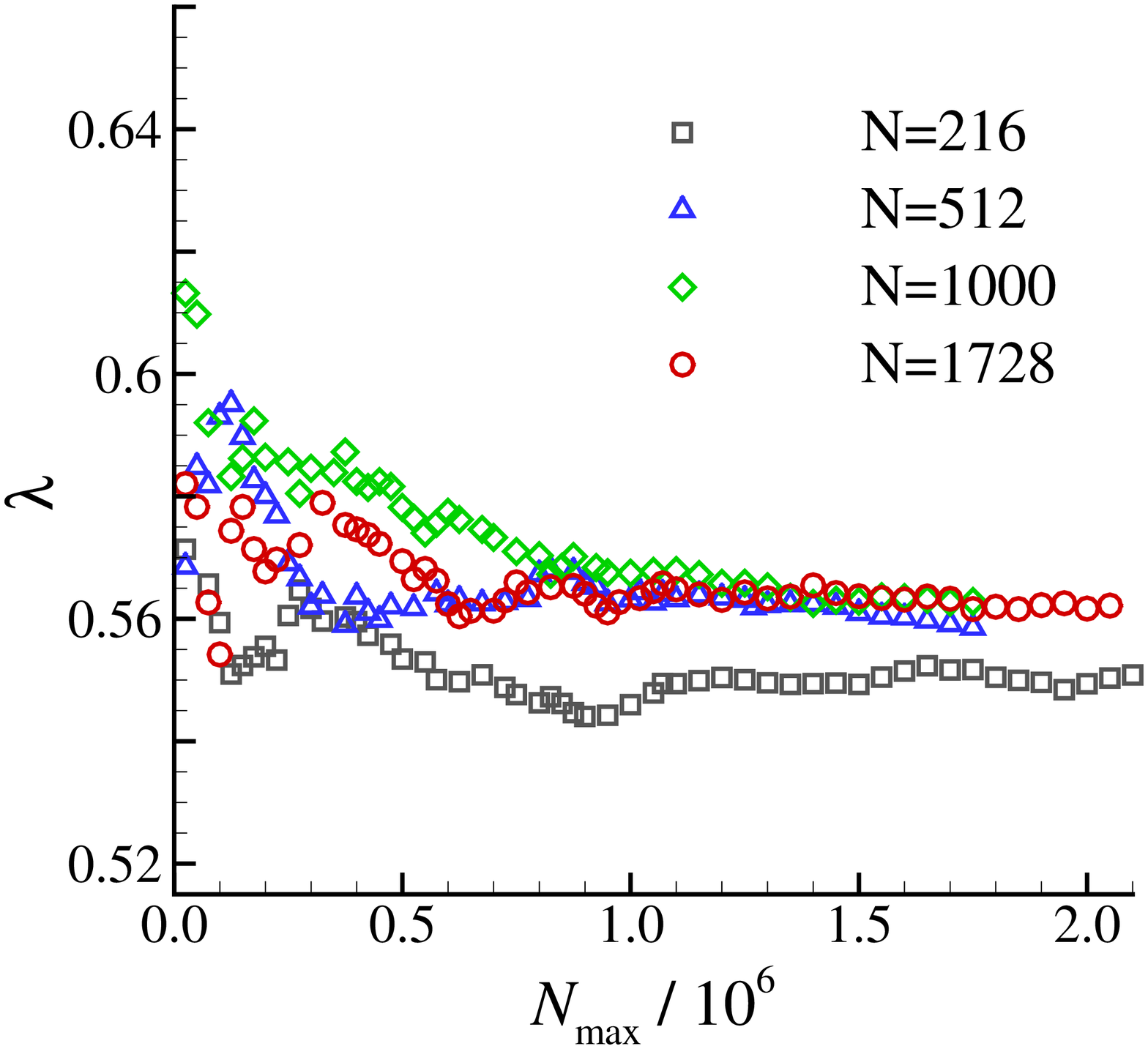}  (c)
		\end{center}\end{minipage} 	\\
		\caption{(color online) Dependence of (a) diffusion coefficient, (b) shear viscosity, and (c) thermal conductivity on system size, $N$, and number of averaging time origins, $N_{max}$, is investigated for the same system as in Fig.\ref{fig:Trans_t} ($\rho=0.018$, $T=1$). For each system size, the value of transport coefficient is calculated at a fixed time step and averaged over $N_{max}$ time origins. By increasing the length of time series (and consequently averaging time origins) the coefficients saturate to a constant value after a transient fluctuating regime. Increasing the system size from $N=512$ to $N=1728$ has only improved the deviations from analytical values by less than one percent. See Sec.\ref{s:R} for more detail.}
		\label{fig:Dep_N}
\end{center}\end{figure*}

A typical measurement for a system of $N=1000$ particles is shown in Fig.~\ref{fig:Trans_t}. As is indicated, transport coefficients, including diffusion coefficient (blue squares), saturate to a constant values after a relatively short transient simulation time ($t\sim200-300$). Although we have chosen a highly relativistic case in low density regime ($T=1,\rho=0.018$), it should be noted that the observed trend is independent of these parameters. The origin of roughnesses observed in tails (long times) is insufficient averaging (low values of $N_{max}$) due to the finite length of time series. To obtain an accurate coefficient, the value of $N_{run}$ is chosen such that a wide plateau forms before the rough zone in the tails. The transport coefficient is then given either by reading one point on the plateau or by averaging on a set of points. Here, and in the following sections, we consider one point on the plateau (dashed line) and increase averaging times ($N_{max}$) such that the finite size effects are retrieved and reliable values for the transport coefficients be obtained (See Fig.~\ref{fig:Dep_N}.).    

\subsection{Shear Viscosity}
Viscosity is a measure of fluid resistivity to deformation that quantifies the strength of friction between layers of fluid. From the viewpoint of transport theory, it characterizes the momentum transport through Newton's law of viscosity, which is read as 
\begin{equation}
\tau_{ij}=\eta_s \frac{\partial u_i}{\partial x_j}
\end{equation}

In 1960, Helfand generalized Einstein relation for self-diffusion and expressed all transport coefficients, in terms of proper fluctuating properties in equilibrium \cite{Hel60}. Shear viscosity $\eta_s$, for example, is given by
\begin{equation}
\eta_s=\frac{1}{2 k_B T V}\lim_{t\rightarrow\infty}\frac{d}{dt}\left\langle |\mathcal {D}_{\alpha\beta}(t)-\mathcal {D}_{\alpha\beta}(0)|^2\right\rangle,
\label{eq:Svisc_Coef}\end{equation}
where 
\begin{equation}
\mathcal {D}_{\alpha\beta}=\sum_{i=1}^{N}r_{i\alpha}p_{i\beta}.
\label{eq:Dab}\end{equation}
In order to numerically determine the shear viscosity, we obtain the quantity  
\begin{eqnarray}
\nonumber &&\left\langle |\mathcal {D}_{\alpha\beta}(t)-\mathcal {D}_{\alpha\beta}(0)|^2\right\rangle=\\
&&\frac{1}{N_{max}}\sum_{n=1}^{N_{max}}|\mathcal {D}_{\alpha\beta}(t_n+t)-\mathcal {D}_{\alpha\beta}(t_n)|^2
\label{eq:Svisc_Coef_Sim}\end{eqnarray}
by measuring $\mathcal {D}_{\alpha\beta}(t)$ at equal time steps and average over $N_{max}$ time origins as discussed in section \ref{ss:SD}. In contrast to the case of diffusion, $\mathcal {D}_{\alpha\beta}$ in this expression explicitly depends on the position of particles, that would cause numerical difficulties when a particle crosses the borders and periodic boundary condition is applied. To overcome this problem, we measure the increment of $\mathcal {D}_{\alpha\beta}$ in well defined time intervals rather than directly measure its value at fixed times. In an event driven molecular dynamics, where the particles move with constant velocity until they collide instantaneously, the increment of $\mathcal {D}_{\alpha\beta}$ in the interval $[t_n,t_n+t]$ can be divided into `kinetic' and `collision' contributions. For time intervals in which no collision occurs, the variation of $\mathcal {D}_{\alpha\beta}$ is purely due to the displacement of the particles 
\begin{equation}
\mathcal {D}_{\alpha\beta}^{(K)}(t_n+\delta \tau)=\sum_{i=1}^{N}\dot r_{i\alpha}(t_n)p_{i\beta} \delta \tau,
\label{eq:Dab_K}\end{equation}
while the discontinuous change of momentum at collisions (between particle $i$ and $j$) gives rise to an instantaneous jump in $\mathcal {D}_{\alpha\beta}$.
\begin{equation}
\mathcal {D}_{\alpha\beta}^{(C)}=r_{i\alpha}\Delta p_{j\beta}+r_{j\alpha}\Delta p_{i\beta},
\label{eq:Dab_C}\end{equation}
with $\Delta p_{j\beta}$ being the change in momentum of particle $j$ in the direction $\beta$. Since $\Delta p_{j\beta}=-\Delta p_{i\beta}$ due to the momentum conservation we can rewrite Eq.~\ref{eq:Dab_C} as
\begin{equation}
\mathcal {D}_{\alpha\beta}^{(C)}=(r_{i\alpha}-r_{j\alpha})\Delta p_{i\beta}.\label{eq:Dab_C2}\end{equation}

With same method, one can measure the bulk viscosity $\eta_b$, using the following expression \cite{All86}:
\begin{equation}
\eta_b+\frac{4}{3}\eta_s=\frac{1}{2 k_B T V}\lim_{t\rightarrow\infty}\frac{d}{dt}\left\langle |\mathcal {D}_{\alpha\alpha}(t)-\mathcal {D}_{\alpha\alpha}(0)-P V t|^2\right\rangle,
\label{eq:Bvisc_Coef}\end{equation}   
with $\mathcal {D}_{\alpha\alpha}=\sum_{i=1}^{N}r_{i\alpha}p_{i\alpha}$, $P$ denoting pressure and $V$ volume of the system. The extra $PVt$ term is the non-vanishing equilibrium average of $\mathcal {D}_{\alpha\alpha}$ which needs to be subtracted \cite{All86}. This quantity is in fact the virial of the system that is evaluated during the simulation. Since it changes as the simulation proceeds, it introduces error to the final value. Another source of error in calculation of bulk viscosity, is the subtraction of shear term $\eta$, which itself has numerical uncertainties. For these reasons, the results obtained for bulk viscosity is very noisy and would not be reported here. Nevertheless, one can follow the same process as discussed for shear viscosity to calculate bulk viscosity. 

\subsection{Thermal Conductivity}
Thermal conductivity characterizes heat transport in a fluid, and is described by the Fourier's law of heat conduction. 
\begin{equation}
J_q=-\lambda_T \nabla T
\end{equation}

In the same line as other transport coefficients, thermal conductivity can also be expressed in Einstein-Helfand form:
\begin{equation}
\lambda_T=\frac{1}{2 k_B T^2 V}\lim_{t\rightarrow\infty}\frac{d}{dt}\left\langle |\delta\varepsilon_{\alpha}(t)-\delta \varepsilon_{\alpha}(0)|^2\right\rangle,
\label{eq:Tcon_Coef}
\end{equation}
with $$\delta\varepsilon_{\alpha}=\sum_{i=1}^{N} r_{i\alpha}(\varepsilon_i-\left\langle h_i\rangle\right).$$ $\varepsilon_i$ is the contribution of $i$-th particle to the total Hamiltonian of the system with the potential energy of interaction of pairs of molecules (if any) being divided equally between partners, $$\varepsilon_i=m_i c^2\gamma(v_i)+\frac{1}{2}\sum_{j\ne i}U(r_{ij}).$$ 
$\langle h_i\rangle=h=K_3(z)/K_2(z)$ is the average enthalpy per molecule. In low energy classical limit, the average enthalpy is often replaced by average energy per particle, $\langle\varepsilon_i\rangle$, without causing any problem. However, it should be noted that such replacement would lead to erroneous result in the relativistic regime. 

The expression to be calculated in simulation is 
\begin{equation}
\left\langle |\delta\varepsilon_{\alpha}(t)-\delta \varepsilon_{\alpha}(0)|^2\right\rangle=\frac{1}{N_{max}}\sum_{n=1}^{N_{max}}|\delta\varepsilon_{\alpha}(t_n+t)-\delta\varepsilon_{\alpha}(t_n)|^2,
\label{eq:Tcon_Coef_Sim}
\end{equation}
which involves the position of particles and becomes problematic as they cross the periodic borders. Here again, the time intervals $[t_n,t_n+t]$ are divided into multiple no-collision intervals ($\delta \tau$) to separately measure the kinetic contribution between collisions,
\begin{equation}
\delta\varepsilon_{\alpha}^{(K)}(t_n+\delta \tau)=\sum_{i=1}^{N}\dot r_{i\alpha}(t_n) (\varepsilon_{i}-\langle h\rangle)\delta \tau,
\label{eq:Ea_K}\end{equation}
and the contribution of instantaneous collisions
\begin{equation}
\delta\varepsilon_{\alpha}^{(C)}=(r_{i\alpha}-r_{j\alpha})\Delta \varepsilon_{i}\label{eq:Ea_C}\end{equation}
For isotropic systems the two contribution in the increment of $\delta\varepsilon$ in a time interval $\delta \tau$ can be rewritten as  
\begin{equation}
\delta\varepsilon^{(K)}=\frac{1}{3}\sum_{\alpha=1}^3\sum_{i=1}^{N}\dot r_{i\alpha} (t_n)(\varepsilon_{i}-\langle h\rangle)\delta \tau
\label{eq:Ea_K_iso}\end{equation}
and
\begin{equation}
\delta\varepsilon^{(C)}=\frac{1}{3}\sum_{\alpha=1}^3(r_{i\alpha}-r_{j\alpha}) \Delta \varepsilon_{i}\label{eq:Ea_C_iso}\end{equation}
 
\section{Results}\label {s:R}

In this section the results of various simulations are presented. Systems with different number of particles and various temperatures are simulated with a fixed density. Measurements are all done in equilibrium state which is reached typically after $10-1000 N$ collisions. To find reliable values with acceptable statistical precision, each quantity is averaged over more than $10^6$ time origins, as discussed in Sec.\ref{s:MST}.

As mentioned before (Fig.~\ref{fig:Trans_t}), the time evolution of transport coefficients saturate to constant values after a relatively short transient simulation time ($t\sim200-300$). Figure \ref{fig:Dep_N} studies the dependence of transport coefficients on $N$ and $N_{max}$. In each subplot, the value of the coefficient is measured at a fixed time step on the plateau (dashed line in Fig.~\ref{fig:Trans_t}), and  plotted versus number of averaging time origins ($N_{max}$) for different system sizes ($N=216, 512, 1000, 1728$). All the systems have identical density $\rho=0.018$  and temperature $T=1$. By increasing the averaging time origins, the coefficients saturate to a constant value after a transient fluctuating regime. The numerical values for $N=1728$ (averaged over values from $N_{max}=(1.5-2)\times 10^6$) give $D=6.534$, $\eta_s=0.396$, and $\lambda=0.555$, which respectively show $8.2, 7.2$, and $1.5$ percent deviation from analytical predictions based on relativistic CE method. Increasing the system size from $N=512$ to $N=1728$ only improves the deviations by less than one percent, so the dependence on system size is relatively negligible for $N \geq 512$. We therefore might safely use $N=512$ and $N_{max}=2\times10^6$ as optimum values from here on.

We now turn to study how transport coefficients vary as a function of temperature. The numerical results would then be used to check the validity of relativistic Chapman-Enskog theory in different temperature regimes. In addition to that, we have performed classical simulations based on Newtonian equations of motion for the particles and compared the resulting transport coefficients with classical predictions \cite{Rei80}. This would provide a check point for testing our simulations, and also illuminates how and at what temperature(s) the relativistic transport coefficients deviate from their classical counterparts.

\begin{figure}
	\centering
	\includegraphics[width=0.8\linewidth]{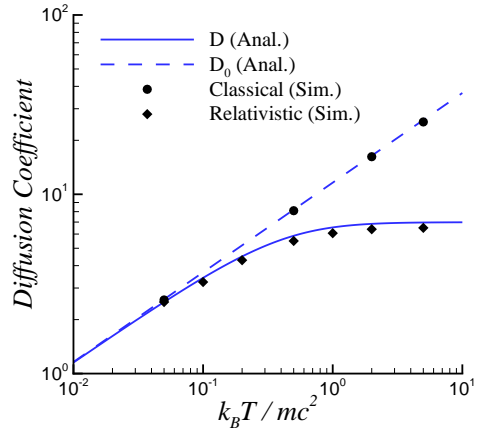}
	\caption{(color online) Diffusion coefficient obtained from classical (circles) and relativistic (diamond) molecular dynamics simulations are compared to their corresponding analytical predictions (dashed and solid lines, respectively). Very good agreement is observed in classical MD (in the whole range of temperature parameters) as well as the low temperature limit of relativistic simulation. Up to $7\%$ deviation from CE theory is observed in highly relativistic limit. Also see Table \ref{t:coeff}.}
	\label{fig:D}
\end{figure}

\begin{figure}
	\centering
	\includegraphics[width=0.8\linewidth]{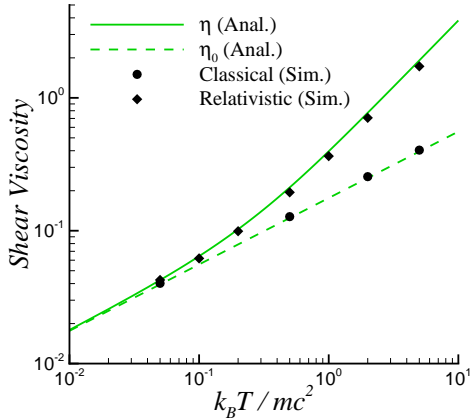}
	\caption{(color online) Shear viscosity obtained from classical (circles) and relativistic (diamond) molecular dynamics simulations are compared to their corresponding analytical predictions (dashed and solid lines, respectively). Very good agreement is observed in classical MD (in the whole range of temperature parameters) as well as the low temperature limit of relativistic simulation. The deviation from theoretical prediction increases from $0.2\%$ at $T=0.05$ to $10\%$ at $T=5$. See Table \ref{t:coeff} for more details.}
	\label{fig:eta}
\end{figure}

\begin{figure}
	\centering
	\includegraphics[width=0.8\linewidth]{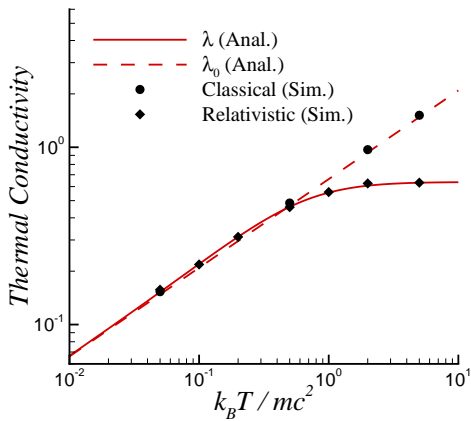}
	\caption{(color online) Thermal conductivity measured from classical (circles) and relativistic (diamond) molecular dynamics simulations are compared to their corresponding analytical predictions (dashed and solid lines, respectively). Classical as well as relativistic MD show very good agreement with theory in the whole range of temperature parameters. Table \ref{t:coeff} gives more detail.}
	\label{fig:lambda}
\end{figure}

Figure \ref{fig:D} indicates the results for self-diffusion transport coefficients as a function of dimensionless parameter $z^{-1}$ for both relativistic and classical systems of $N=512$ and $\rho=0.018$. Dashed and solid lines are respectively the analytical predictions from classical and relativistic theories. It is observed that the numerical values of self-diffusion coefficient are in good agreement with analytical predictions for classical systems ($D_0=3/(32 \rho R^2)\sqrt{k_B T/m\pi}$) as well as low temperature limit of relativistic theory. In high temperature limit, however, the results of relativistic MD show up to $7\%$ deviation from CE prediction (Eq.\ref{e:CE_D}). Since the system is in low density regime, the observed deviation might be asserted to the inaccuracy of relativistic CE prediction in high temperature regime.

Same behavior is observed in Figure \ref{fig:eta}, where shear viscosity of the system is presented. As is evident, the classical simulation agrees very well with analytical prediction ($\eta_0=5/(64 R^2)\sqrt{mk_B T/\pi}$), while the deviation of numerical results from CE prediction increases as temperature rises. Increasing the system size from $N=512$ to $N=1728$ hardly improves the results (about one percent). It can therefore be concluded that the monotonic increase of deviations from $0.2\%$ at $T=0.05$ to $10\%$ at $T=5$ suggests the inaccuracy of CE approximation in high temperature regime rather than system size effects.  

In the case of thermal conductivity, shown in Fig.~\ref{fig:lambda}, not only the  classical theory ($\lambda_0=75/(256 R^2)\sqrt{k_B^3 T/m\pi}$) agrees with numerical results, but also a very good agreement is found between relativistic CE prediction and simulation data, in the whole range of temperature. This is in contrast to the temperature dependent deviations observed for the other two coefficient summarized in Table~\ref{t:coeff}. 

Another interesting issue is the behavior of classical coefficients (dashed lines in Figs.~\ref{fig:D} to \ref{fig:lambda}) versus their relativistic counterpart (solid lines), and the question of how and at what temperature the relativistic effects would emerge. 

Unlike the unbounded linear increase in classical theory, the relativistic diffusion coefficient as well as thermal conductivity lead to a saturating behavior for $T\gtrsim1$, which is due to the existence of an upper limit for velocity of particles and consequently for propagation of non-uniformities in diffusive phenomena. The relativistic shear viscosity coefficient, on the other hand, is always larger than the classical one, indicating that the relativistic fluid is  more resistive to deformations and shows larger interlayer friction. 

\begin{figure}
	\centering
	\includegraphics[width=0.8\linewidth]{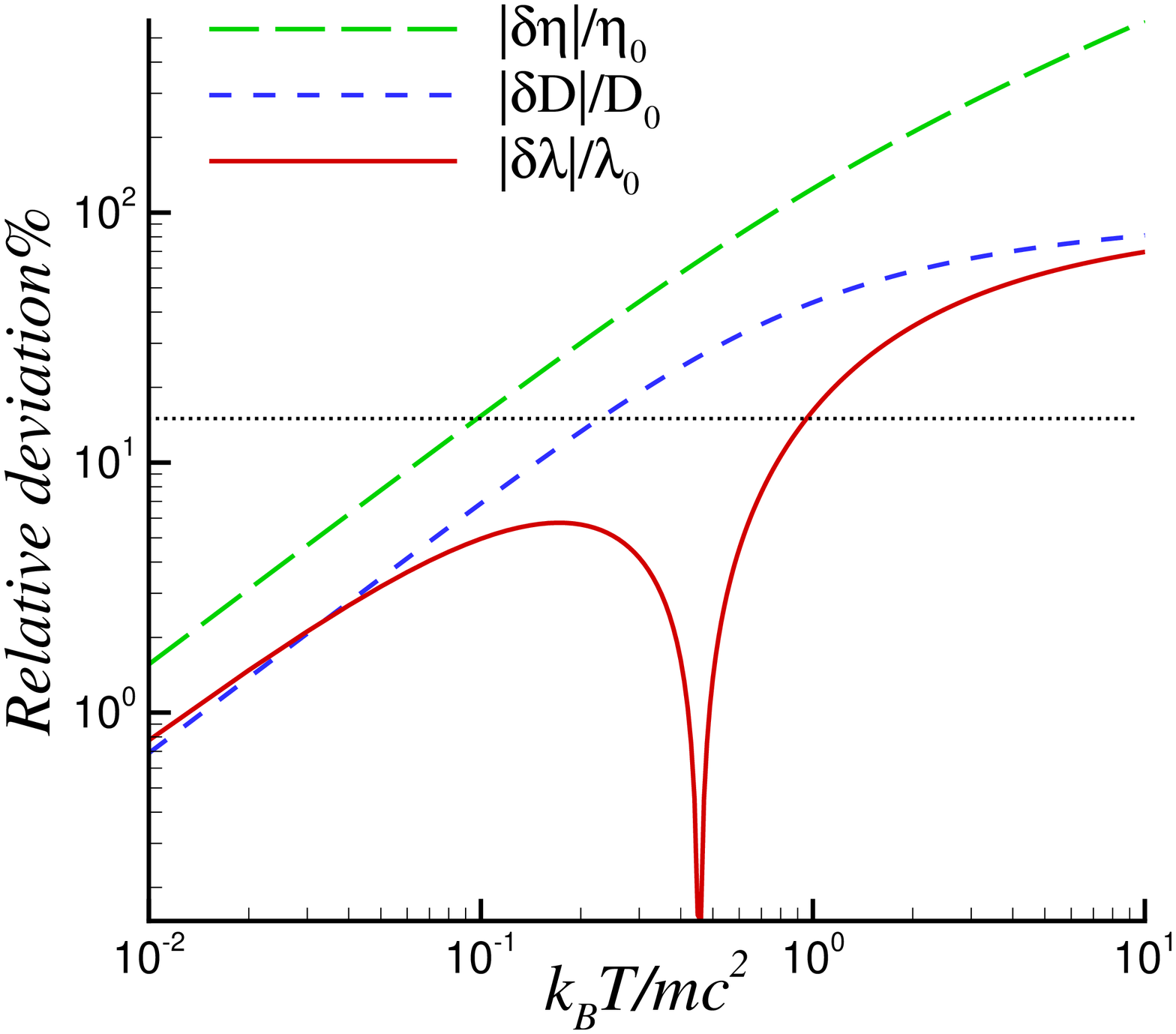}
	\caption{(color online) Relative deviation of relativistic coefficients compared to their classical counterpart is plotted versus dimensionless temperature parameter. A deviation of $15$ percent (dotted line), which we have defined as the emergence of relativistic effects, is observed at $T\simeq0.1$, $T\simeq0.25$ and $T\simeq1$ for viscosity (long dashed green line), diffusion coefficient (dashed blue line) and thermal conductivity (solid red line), respectively.}
	\label{fig:diffperc}
\end{figure}

Figure~\ref{fig:diffperc} studies the relative deviation of relativistic coefficients from their classical counterpart as a function of temperature. Defining the beginning of a notable deviation (which is set $\sim15$ percent here) as the emergence of relativistic effects in transport coefficient, one observes that it occurs around $T\simeq0.25$ and $T\simeq0.1$ for diffusion (dashed blue line) and shear viscosity (long dashed green line), respectively. In the case of thermal conductivity (solid red line), the classical prediction is almost equal to its relativistic counterpart for temperatures up to $T\simeq0.5$, and relativistic effects are only found for $T\gtrsim1$. 

\begin{table}[t]
	\begin{tabular}{cccc}
		\hline
		\hline
		$T$ & \hspace{10mm} $D_{sim}/D_{th} $  & \hspace{10mm}  $\eta_{sim}/\eta_{th}$ & \hspace{10mm} $\lambda_{sim}/\lambda_{th}$\\
		\hline
		0.05& \hspace{10mm} 0.97 & \hspace{10mm} 1.00   & \hspace{10mm} 1.03  \\
		0.1 & \hspace{10mm} 0.95 & \hspace{10mm} 0.96   & \hspace{10mm} 0.99  \\
		0.2 & \hspace{10mm} 0.95 & \hspace{10mm} 0.97   & \hspace{10mm} 1.00  \\
		0.5 & \hspace{10mm} 0.93 & \hspace{10mm} 0.92   & \hspace{10mm} 1.00  \\
		1.0 & \hspace{10mm} 0.93 & \hspace{10mm} 0.92   & \hspace{10mm} 1.01  \\
		2.0 & \hspace{10mm} 0.93 & \hspace{10mm} 0.91   & \hspace{10mm} 1.03  \\
		5.0 & \hspace{10mm} 0.93 & \hspace{10mm} 0.90   & \hspace{10mm} 1.00	 \\
		\hline	
		\hline	
	\end{tabular}
	\caption{Ratio of simulation data to analytical values at different temperatures $T$, is given for a system of $N=512$ and $\rho=0.018$. As temperature rises deviations in diffusion coefficient and shear viscosity show an increase, while the errors in thermal conductivity is independent of temperature.}
	\label{t:coeff}
\end{table}

Since any macroscopic phenomenon is 
a manifestation of the underlying microscopic properties, it is worth recalling the behavior of equilibrium velocity distribution of a  relativistic gas (as a microscopic property) in comparison with the emergence of relativistic effects in transport coefficients (as macroscopic properties). It has been shown that, the single-peaked Maxwell-Boltzmann distribution undergoes a phase transition to a double-peaked J\"{u}ttner distribution at the critical temperature $T_c=1/(d+2)$ with $d$ being the system dimension \cite{Men13,Mon14}. This is considered to be the transition from classical to relativistic regime. For a three dimensional case, the critical temperature ($T_c=0.2$) is very close to the value where notable deviation of relativistic diffusion from its classical counterpart is observed. The deviations in viscosity (thermal conductivity) is larger (smaller) compared to diffusion and occurs earlier (later); i.e. in lower (higher) temperatures as is seen in figure~\ref{fig:diffperc}.   

Comparing the results summarized in table~\ref{t:coeff} and figure~\ref{fig:diffperc}, it is inferred that the earlier the relativistic effects emerge in a coefficient, the more deviation of simulation data from CE theory would be seen; such that the largest deviation and best agreement are respectively observed in viscosity and thermal conductivity. This completes our primitive conclusion that the deviation of numerical data from theory is due to the insufficiency of first order CE theory in high temperature limit where relativistic effects are most strongly expressed. The current data, however, does not justify why relativistic effect emerges earlier and/or later in one coefficient compared to the other.  

\section{Conclusion}\label{s:C}
In this work the Einstein-Helfand relations has been successfully employed to numerically obtain the relativistic transport coefficients of a massive hard sphere gas is measured based on an event driven molecular dynamics simulation. Since, the model is a fully relativistic one, it would in principle provide numerically exact results, which in the absence of experimental data, is the only reliable tool for investigating the accuracy of relativistic theories for various temperature and density regimes. 

The systems under study in this manuscript are all in low density regime, where the linear CE theory is believed to be a good description, while temperature parameter covers a wide range, from low temperature classical regime to extremely relativistic limit. According to the results summarized in Table 1, it can generally be concluded that the relativistic CE theory (as a first-order theory) gives relatively reliable values for transport coefficients of a low density hard sphere gaseous system.

In a closer look, however, deviations are observed from CE analytical predictions (up to 10 percent in some extremely relativistic cases). Since the simulations are conducted in low enough density regimes, and the observed deviations (especially in diffusion coefficient and shear viscosity) increase by rising the temperature, the discrepancies could be well attributed to `inaccuracy of CE approximation' in highly relativistic regime rather than high density or system size impacts as discussed in Fig.~\ref{fig:Dep_N}.

Another interesting result is that thermal conductivity, in contrast to the other two coefficients, shows a very good agreement with CE theory in the whole range of temperature parameters (less than 3 percent deviation). Such a difference is also observed between propagation of thermal and acoustic modes (which are respectively controlled by thermal conductivity and shear viscosity) in a relativistic fluid \cite{Gho16}. These evidences in addition to the fact that relativistic effects in thermal conductivity emerges at higher temperatures compared to $D$ and $\eta$ (Fig.~\ref{fig:diffperc}), suggest that there is a negative correlation between the emergence of relativistic effects in each coefficient and accuracy of CE prediction in high temperature limit; such that earlier emergence of these effects corresponds to higher deviation of data from CE theory. However, the reason why relativistic effects emerge earlier in one coefficient compared to the other remains unanswered based on the current numerical measurements. More accurate simulations, specifically around the transition points where relativistic effects begin to emerge and extreme cases where they are fully expressed might be helpful to shed light on the issue.  

Finally, it should be noted that the reported findings are for non-degenerate gases. The generalization to Fermi-Dirac and Bose-Einstein statistics is straight forward and does not affect the CE approximation or accuracy of transport coefficient \cite{Gro80,Cer02}. Therefore, it can be inferred that the main results remain unchanged for degenerate systems.  

\section*{Acknowledgments}
The authors are grateful to ..... for useful discussions/comments. This work is supported by Tarbiat Modares University Research Council.  

\bibliographystyle{apsrev}
\bibliography{xbib}

\begin{thebibliography}{46}
\expandafter\ifx\csname natexlab\endcsname\relax\def\natexlab#1{#1}\fi
\expandafter\ifx\csname bibnamefont\endcsname\relax
  \def\bibnamefont#1{#1}\fi
\expandafter\ifx\csname bibfnamefont\endcsname\relax
  \def\bibfnamefont#1{#1}\fi
\expandafter\ifx\csname citenamefont\endcsname\relax
  \def\citenamefont#1{#1}\fi
\expandafter\ifx\csname url\endcsname\relax
  \def\url#1{\texttt{#1}}\fi
\expandafter\ifx\csname urlprefix\endcsname\relax\def\urlprefix{URL }\fi
\providecommand{\bibinfo}[2]{#2}
\providecommand{\eprint}[2][]{\url{#2}}

\bibitem[{\citenamefont{Reichl}(1998)}]{Rei80}
\bibinfo{author}{\bibfnamefont{L.~E.} \bibnamefont{Reichl}},
  \emph{\bibinfo{title}{A Modern Course in Statistical Physics}}
  (\bibinfo{publisher}{John Wiley and Sons}, \bibinfo{address}{N. Y.},
  \bibinfo{year}{1998}), \bibinfo{edition}{2nd} ed.

\bibitem[{\citenamefont{J\"uttner}(1911)}]{Jut11}
\bibinfo{author}{\bibfnamefont{F.}~\bibnamefont{J\"uttner}},
  \bibinfo{journal}{Ann. Phys.} \textbf{\bibinfo{volume}{339}},
  \bibinfo{pages}{856} (\bibinfo{year}{1911}).

\bibitem[{\citenamefont{Licherowicz and Marrot}(1940)}]{Lich40}
\bibinfo{author}{\bibfnamefont{A.}~\bibnamefont{Licherowicz}} \bibnamefont{and}
  \bibinfo{author}{\bibfnamefont{R.}~\bibnamefont{Marrot}},
  \bibinfo{journal}{Comp. Rend. Acad. Sci. (Paris)}
  \textbf{\bibinfo{volume}{210}}, \bibinfo{pages}{759} (\bibinfo{year}{1940}).

\bibitem[{\citenamefont{Israel}(1963)}]{Isr63}
\bibinfo{author}{\bibfnamefont{W.}~\bibnamefont{Israel}}, \bibinfo{journal}{J.
  Math. Phys.} \textbf{\bibinfo{volume}{4}}, \bibinfo{pages}{1163}
  (\bibinfo{year}{1963}).

\bibitem[{\citenamefont{Kelly}(1963)}]{Kel63}
\bibinfo{author}{\bibfnamefont{D.~C.} \bibnamefont{Kelly}},
  \emph{\bibinfo{title}{The Kinetic Theory of a Relativistic Gas}}
  (\bibinfo{publisher}{Miami Univ.}, \bibinfo{address}{Oxford},
  \bibinfo{year}{1963}).

\bibitem[{\citenamefont{Chernikov}(1964)}]{Cher64}
\bibinfo{author}{\bibfnamefont{N.~A.} \bibnamefont{Chernikov}},
  \bibinfo{journal}{Act. Phys. Pol.} \textbf{\bibinfo{volume}{27}},
  \bibinfo{pages}{465} (\bibinfo{year}{1964}).

\bibitem[{\citenamefont{de~Groot et~al.}(1980)\citenamefont{de~Groot, van
  Leeuwen, and van Weert}}]{Gro80}
\bibinfo{author}{\bibfnamefont{S.~R.} \bibnamefont{de~Groot}},
  \bibinfo{author}{\bibfnamefont{W.~A.} \bibnamefont{van Leeuwen}},
  \bibnamefont{and} \bibinfo{author}{\bibfnamefont{C.~G.} \bibnamefont{van
  Weert}}, \emph{\bibinfo{title}{Relativistic Kinetic Theory: Principles and
  Applications}} (\bibinfo{publisher}{North-Holand},
  \bibinfo{address}{Amsterdam}, \bibinfo{year}{1980}).

\bibitem[{\citenamefont{Cercignani and Kremer}(2002)}]{Cer02}
\bibinfo{author}{\bibfnamefont{C.}~\bibnamefont{Cercignani}} \bibnamefont{and}
  \bibinfo{author}{\bibfnamefont{G.~M.} \bibnamefont{Kremer}},
  \emph{\bibinfo{title}{The Relativistic Boltzmann Equation: Theory and
  Applications}}, vol.~\bibinfo{volume}{22} of \emph{\bibinfo{series}{Progress
  in Mathematical Physics}} (\bibinfo{publisher}{Birkh\"{a}user Verlag},
  \bibinfo{address}{Berlin}, \bibinfo{year}{2002}).

\bibitem[{\citenamefont{Rezzolla and Zanotti}(2013)}]{Rez13}
\bibinfo{author}{\bibfnamefont{L.}~\bibnamefont{Rezzolla}} \bibnamefont{and}
  \bibinfo{author}{\bibfnamefont{O.}~\bibnamefont{Zanotti}},
  \emph{\bibinfo{title}{Relativistic Hydrodynamics}}
  (\bibinfo{publisher}{Oxford Univ. Press}, \bibinfo{address}{UK},
  \bibinfo{year}{2013}).

\bibitem[{\citenamefont{Lopez-Monsalvo and Andersson}(2011)}]{Lop11}
\bibinfo{author}{\bibfnamefont{C.~S.} \bibnamefont{Lopez-Monsalvo}}
  \bibnamefont{and}
  \bibinfo{author}{\bibfnamefont{N.}~\bibnamefont{Andersson}},
  \bibinfo{journal}{Proc. Roy. Soc. London A} \textbf{\bibinfo{volume}{467}},
  \bibinfo{pages}{738} (\bibinfo{year}{2011}).

\bibitem[{\citenamefont{Balsara}(2001)}]{Bal01}
\bibinfo{author}{\bibfnamefont{D.}~\bibnamefont{Balsara}},
  \bibinfo{journal}{Astrophys. J. Suppl.} \textbf{\bibinfo{volume}{132}},
  \bibinfo{pages}{83} (\bibinfo{year}{2001}).

\bibitem[{\citenamefont{Huovinen}(2004)}]{Huo04}
\bibinfo{author}{\bibfnamefont{P.}~\bibnamefont{Huovinen}}, in
  \emph{\bibinfo{booktitle}{Quark Gluon Plasma 3}}, edited by
  \bibinfo{editor}{\bibfnamefont{R.~C.} \bibnamefont{Hwa}} \bibnamefont{and}
  \bibinfo{editor}{\bibfnamefont{X.~N.} \bibnamefont{Wang}}
  (\bibinfo{publisher}{Word Scientific}, \bibinfo{address}{Singapore},
  \bibinfo{year}{2004}), p. \bibinfo{pages}{600}.

\bibitem[{\citenamefont{Kolb and Heinz}(2004)}]{Kol04}
\bibinfo{author}{\bibfnamefont{P.~F.} \bibnamefont{Kolb}} \bibnamefont{and}
  \bibinfo{author}{\bibfnamefont{U.}~\bibnamefont{Heinz}}, in
  \emph{\bibinfo{booktitle}{Quark Gluon Plasma 3}}, edited by
  \bibinfo{editor}{\bibfnamefont{R.~C.} \bibnamefont{Hwa}} \bibnamefont{and}
  \bibinfo{editor}{\bibfnamefont{X.~N.} \bibnamefont{Wang}}
  (\bibinfo{publisher}{Word Scientific}, \bibinfo{address}{Singapore},
  \bibinfo{year}{2004}), p. \bibinfo{pages}{634}.

\bibitem[{\citenamefont{M\"{u}ller and Sachdev}(2008)}]{Mul08a}
\bibinfo{author}{\bibfnamefont{M.}~\bibnamefont{M\"{u}ller}} \bibnamefont{and}
  \bibinfo{author}{\bibfnamefont{S.}~\bibnamefont{Sachdev}},
  \bibinfo{journal}{Phys. Rev. B} \textbf{\bibinfo{volume}{78}},
  \bibinfo{pages}{115419} (\bibinfo{year}{2008}).

\bibitem[{\citenamefont{M\"{u}ller et~al.}(2009)\citenamefont{M\"{u}ller,
  Schmalian, and Fritz}}]{Mul09}
\bibinfo{author}{\bibfnamefont{M.}~\bibnamefont{M\"{u}ller}},
  \bibinfo{author}{\bibfnamefont{J.}~\bibnamefont{Schmalian}},
  \bibnamefont{and} \bibinfo{author}{\bibfnamefont{L.}~\bibnamefont{Fritz}},
  \bibinfo{journal}{Phys. Rev. Lett.} \textbf{\bibinfo{volume}{103}},
  \bibinfo{pages}{025301} (\bibinfo{year}{2009}).

\bibitem[{\citenamefont{Mendoza et~al.}(2011)\citenamefont{Mendoza, Herrmann,
  and Succi}}]{Men11}
\bibinfo{author}{\bibfnamefont{M.}~\bibnamefont{Mendoza}},
  \bibinfo{author}{\bibfnamefont{H.~J.} \bibnamefont{Herrmann}},
  \bibnamefont{and} \bibinfo{author}{\bibfnamefont{S.}~\bibnamefont{Succi}},
  \bibinfo{journal}{Phys. Rev. Lett.} \textbf{\bibinfo{volume}{106}},
  \bibinfo{pages}{156601} (\bibinfo{year}{2011}).

\bibitem[{\citenamefont{Torre et~al.}(2015)\citenamefont{Torre, Tomadin, Geim,
  and Polini}}]{Tor15}
\bibinfo{author}{\bibfnamefont{I.}~\bibnamefont{Torre}},
  \bibinfo{author}{\bibfnamefont{A.}~\bibnamefont{Tomadin}},
  \bibinfo{author}{\bibfnamefont{A.~K.} \bibnamefont{Geim}}, \bibnamefont{and}
  \bibinfo{author}{\bibfnamefont{M.}~\bibnamefont{Polini}},
  \bibinfo{journal}{Phys. Rev. B} \textbf{\bibinfo{volume}{92}},
  \bibinfo{pages}{165433} (\bibinfo{year}{2015}).

\bibitem[{\citenamefont{Israel}(1976)}]{Isr76}
\bibinfo{author}{\bibfnamefont{W.}~\bibnamefont{Israel}},
  \bibinfo{journal}{Ann. Phys} \textbf{\bibinfo{volume}{100}},
  \bibinfo{pages}{310} (\bibinfo{year}{1976}).

\bibitem[{\citenamefont{Stewart}(1977)}]{Stew77}
\bibinfo{author}{\bibfnamefont{J.~M.} \bibnamefont{Stewart}},
  \bibinfo{journal}{Roy. Soc. London Proc. Series A}
  \textbf{\bibinfo{volume}{357}}, \bibinfo{pages}{59} (\bibinfo{year}{1977}).

\bibitem[{\citenamefont{Liu et~al.}(1986)\citenamefont{Liu, M\"{u}ller, and
  Ruggeri}}]{Liu86}
\bibinfo{author}{\bibfnamefont{I.~S.} \bibnamefont{Liu}},
  \bibinfo{author}{\bibfnamefont{I.}~\bibnamefont{M\"{u}ller}},
  \bibnamefont{and} \bibinfo{author}{\bibfnamefont{T.}~\bibnamefont{Ruggeri}},
  \bibinfo{journal}{Ann. Phys.} \textbf{\bibinfo{volume}{169}},
  \bibinfo{pages}{191} (\bibinfo{year}{1986}).

\bibitem[{\citenamefont{M\"{u}ller}(1967)}]{Muel67}
\bibinfo{author}{\bibfnamefont{I.}~\bibnamefont{M\"{u}ller}},
  \bibinfo{journal}{Zeitschrift f\"{u}r Physik} \textbf{\bibinfo{volume}{198}},
  \bibinfo{pages}{329} (\bibinfo{year}{1967}).

\bibitem[{\citenamefont{Liu and M\"{u}ller}(1983)}]{Liu83}
\bibinfo{author}{\bibfnamefont{I.~S.} \bibnamefont{Liu}} \bibnamefont{and}
  \bibinfo{author}{\bibfnamefont{I.}~\bibnamefont{M\"{u}ller}},
  \bibinfo{journal}{Arch. Ration. Mech. An.} \textbf{\bibinfo{volume}{83}},
  \bibinfo{pages}{285} (\bibinfo{year}{1983}).

\bibitem[{\citenamefont{Garc\'{\i}a-Col\'{\i}n and
  Sandoval-Villalbazo}(2006)}]{Gcol06}
\bibinfo{author}{\bibfnamefont{L.~S.} \bibnamefont{Garc\'{\i}a-Col\'{\i}n}}
  \bibnamefont{and}
  \bibinfo{author}{\bibfnamefont{A.}~\bibnamefont{Sandoval-Villalbazo}},
  \bibinfo{journal}{J. Nonequil. Therm.} \textbf{\bibinfo{volume}{31}},
  \bibinfo{pages}{11} (\bibinfo{year}{2006}), \bibinfo{note}{also see: A. L.
  Garc\'{\i}a-Perciante, A. Sandoval-Villalbazo, J. Non-Newtonian Fluid Mech.
  165 (2010) 1024}.

\bibitem[{\citenamefont{Garc\'{\i}a-Perciante
  et~al.}(2009)\citenamefont{Garc\'{\i}a-Perciante, Garc\'{\i}a-Col\'{\i}n, and
  Sandoval-Villalbazo}}]{Gper09b}
\bibinfo{author}{\bibfnamefont{A.~L.} \bibnamefont{Garc\'{\i}a-Perciante}},
  \bibinfo{author}{\bibfnamefont{L.~S.} \bibnamefont{Garc\'{\i}a-Col\'{\i}n}},
  \bibnamefont{and}
  \bibinfo{author}{\bibfnamefont{A.}~\bibnamefont{Sandoval-Villalbazo}},
  \bibinfo{journal}{Phys. Rev. E} \textbf{\bibinfo{volume}{79}},
  \bibinfo{pages}{066310} (\bibinfo{year}{2009}).

\bibitem[{\citenamefont{Garc\'{\i}a-Perciante and
  Sandoval-Villalbazo}(2010)}]{Gper10}
\bibinfo{author}{\bibfnamefont{A.~L.} \bibnamefont{Garc\'{\i}a-Perciante}}
  \bibnamefont{and}
  \bibinfo{author}{\bibfnamefont{A.}~\bibnamefont{Sandoval-Villalbazo}},
  \bibinfo{journal}{J. Non-Newtonian Fluid Mech.}
  \textbf{\bibinfo{volume}{165}}, \bibinfo{pages}{1024} (\bibinfo{year}{2010}).

\bibitem[{\citenamefont{Tsumura et~al.}(2007)\citenamefont{Tsumura, Kunihiro,
  and Ohnishi}}]{Tsum07}
\bibinfo{author}{\bibfnamefont{K.}~\bibnamefont{Tsumura}},
  \bibinfo{author}{\bibfnamefont{T.}~\bibnamefont{Kunihiro}}, \bibnamefont{and}
  \bibinfo{author}{\bibfnamefont{K.}~\bibnamefont{Ohnishi}},
  \bibinfo{journal}{Phys. Lett. B.} \textbf{\bibinfo{volume}{646}},
  \bibinfo{pages}{134} (\bibinfo{year}{2007}).

\bibitem[{\citenamefont{V\'{a}n and Bir\'{o}}(2012)}]{Van12}
\bibinfo{author}{\bibfnamefont{P.}~\bibnamefont{V\'{a}n}} \bibnamefont{and}
  \bibinfo{author}{\bibfnamefont{T.~S.} \bibnamefont{Bir\'{o}}},
  \bibinfo{journal}{Phys. Lett. B} \textbf{\bibinfo{volume}{709}},
  \bibinfo{pages}{106} (\bibinfo{year}{2012}).

\bibitem[{\citenamefont{Ghodrat and Montakhab}(2013)}]{Gho13}
\bibinfo{author}{\bibfnamefont{M.}~\bibnamefont{Ghodrat}} \bibnamefont{and}
  \bibinfo{author}{\bibfnamefont{A.}~\bibnamefont{Montakhab}},
  \bibinfo{journal}{Phys. Rev. E.} \textbf{\bibinfo{volume}{87}},
  \bibinfo{pages}{032120} (\bibinfo{year}{2013}).

\bibitem[{\citenamefont{Shahsavar et~al.}(2016)\citenamefont{Shahsavar,
  Ghodrat, and Montakhab}}]{Gho16}
\bibinfo{author}{\bibfnamefont{L.}~\bibnamefont{Shahsavar}},
  \bibinfo{author}{\bibfnamefont{M.}~\bibnamefont{Ghodrat}}, \bibnamefont{and}
  \bibinfo{author}{\bibfnamefont{A.}~\bibnamefont{Montakhab}},
  \bibinfo{journal}{Phys. Rev. C.} \textbf{\bibinfo{volume}{94}},
  \bibinfo{pages}{064905} (\bibinfo{year}{2016}).

\bibitem[{\citenamefont{Green}(1951)}]{Gre51}
\bibinfo{author}{\bibfnamefont{M.~S.} \bibnamefont{Green}},
  \bibinfo{journal}{J. Chem. Phys.} \textbf{\bibinfo{volume}{19}},
  \bibinfo{pages}{1036} (\bibinfo{year}{1951}).

\bibitem[{\citenamefont{Kubo}(1957)}]{Kub57}
\bibinfo{author}{\bibfnamefont{R.}~\bibnamefont{Kubo}}, \bibinfo{journal}{J.
  Phys. Soc. Jpn.} \textbf{\bibinfo{volume}{12}}, \bibinfo{pages}{570}
  (\bibinfo{year}{1957}).

\bibitem[{\citenamefont{Alder et~al.}(1970)\citenamefont{Alder, Gass, and
  Wainwright}}]{Ald70}
\bibinfo{author}{\bibfnamefont{B.~J.} \bibnamefont{Alder}},
  \bibinfo{author}{\bibfnamefont{D.~M.} \bibnamefont{Gass}}, \bibnamefont{and}
  \bibinfo{author}{\bibfnamefont{T.~E.} \bibnamefont{Wainwright}},
  \bibinfo{journal}{J. Chem. Phys.} \textbf{\bibinfo{volume}{53}},
  \bibinfo{pages}{3813} (\bibinfo{year}{1970}).

\bibitem[{\citenamefont{Helfand}(1960)}]{Hel60}
\bibinfo{author}{\bibfnamefont{E.}~\bibnamefont{Helfand}},
  \bibinfo{journal}{Phys. Rev.} \textbf{\bibinfo{volume}{119}},
  \bibinfo{pages}{1} (\bibinfo{year}{1960}).

\bibitem[{\citenamefont{Garc\'{i}a-Rojo
  et~al.}(2006)\citenamefont{Garc\'{i}a-Rojo, Luding, and Brey}}]{Gar06}
\bibinfo{author}{\bibfnamefont{R.}~\bibnamefont{Garc\'{i}a-Rojo}},
  \bibinfo{author}{\bibfnamefont{S.}~\bibnamefont{Luding}}, \bibnamefont{and}
  \bibinfo{author}{\bibfnamefont{J.~J.} \bibnamefont{Brey}},
  \bibinfo{journal}{Phys. Rev. E} \textbf{\bibinfo{volume}{74}},
  \bibinfo{pages}{061305} (\bibinfo{year}{2006}).

\bibitem[{\citenamefont{Viscardy et~al.}(2007)\citenamefont{Viscardy,
  Servantie, and Gaspard}}]{Vis07}
\bibinfo{author}{\bibfnamefont{S.}~\bibnamefont{Viscardy}},
  \bibinfo{author}{\bibfnamefont{J.}~\bibnamefont{Servantie}},
  \bibnamefont{and} \bibinfo{author}{\bibfnamefont{P.}~\bibnamefont{Gaspard}},
  \bibinfo{journal}{J. Chem. Phys.} \textbf{\bibinfo{volume}{126}},
  \bibinfo{pages}{184512} (\bibinfo{year}{2007}).

\bibitem[{\citenamefont{Chapman}(1916)}]{Chap16}
\bibinfo{author}{\bibfnamefont{S.}~\bibnamefont{Chapman}},
  \bibinfo{journal}{Phil. Trans. Roy. London} \textbf{\bibinfo{volume}{216}},
  \bibinfo{pages}{279} (\bibinfo{year}{1916}).

\bibitem[{\citenamefont{Grad}(1949)}]{Gra49}
\bibinfo{author}{\bibfnamefont{H.}~\bibnamefont{Grad}}, \bibinfo{journal}{Comm.
  Pure Appl. Math} \textbf{\bibinfo{volume}{2}}, \bibinfo{pages}{331}
  (\bibinfo{year}{1949}).

\bibitem[{\citenamefont{Cercignani and Kremer}(2001)}]{Cer01}
\bibinfo{author}{\bibfnamefont{C.}~\bibnamefont{Cercignani}} \bibnamefont{and}
  \bibinfo{author}{\bibfnamefont{G.~M.} \bibnamefont{Kremer}},
  \bibinfo{journal}{Physica A} \textbf{\bibinfo{volume}{290}},
  \bibinfo{pages}{192} (\bibinfo{year}{2001}).

\bibitem[{\citenamefont{Anderson and Witting}(1974)}]{And74}
\bibinfo{author}{\bibfnamefont{J.~L.} \bibnamefont{Anderson}} \bibnamefont{and}
  \bibinfo{author}{\bibfnamefont{H.~R.} \bibnamefont{Witting}},
  \bibinfo{journal}{Physica} \textbf{\bibinfo{volume}{74}},
  \bibinfo{pages}{466} (\bibinfo{year}{1974}).

\bibitem[{\citenamefont{Callen and Welton}(1951)}]{Cal51}
\bibinfo{author}{\bibfnamefont{H.~B.} \bibnamefont{Callen}} \bibnamefont{and}
  \bibinfo{author}{\bibfnamefont{T.~A.} \bibnamefont{Welton}},
  \bibinfo{journal}{Phys. Rev.} \textbf{\bibinfo{volume}{83}},
  \bibinfo{pages}{34} (\bibinfo{year}{1951}).

\bibitem[{\citenamefont{Ashurst and Hoover}(1973)}]{Ash73}
\bibinfo{author}{\bibfnamefont{W.~T.} \bibnamefont{Ashurst}} \bibnamefont{and}
  \bibinfo{author}{\bibfnamefont{W.~G.} \bibnamefont{Hoover}},
  \bibinfo{journal}{Phys. Rev. Lett.} \textbf{\bibinfo{volume}{31}},
  \bibinfo{pages}{206} (\bibinfo{year}{1973}).

\bibitem[{\citenamefont{Allen and Tildesley}(1986)}]{All86}
\bibinfo{author}{\bibfnamefont{M.~P.} \bibnamefont{Allen}} \bibnamefont{and}
  \bibinfo{author}{\bibfnamefont{D.~J.} \bibnamefont{Tildesley}},
  \emph{\bibinfo{title}{Computer Simulation of Liquids}}
  (\bibinfo{publisher}{Oxford Univ. Press}, \bibinfo{address}{N. Y.},
  \bibinfo{year}{1986}).

\bibitem[{\citenamefont{Ghodrat and Montakhab}(2011)}]{Gho11}
\bibinfo{author}{\bibfnamefont{M.}~\bibnamefont{Ghodrat}} \bibnamefont{and}
  \bibinfo{author}{\bibfnamefont{A.}~\bibnamefont{Montakhab}},
  \bibinfo{journal}{Com. Phys. Comm.} \textbf{\bibinfo{volume}{182}},
  \bibinfo{pages}{1909} (\bibinfo{year}{2011}).

\bibitem[{\citenamefont{Montakhab et~al.}(2009)\citenamefont{Montakhab,
  Ghodrat, and Barati}}]{Mon09}
\bibinfo{author}{\bibfnamefont{A.}~\bibnamefont{Montakhab}},
  \bibinfo{author}{\bibfnamefont{M.}~\bibnamefont{Ghodrat}}, \bibnamefont{and}
  \bibinfo{author}{\bibfnamefont{M.}~\bibnamefont{Barati}},
  \bibinfo{journal}{Phys. Rev. E} \textbf{\bibinfo{volume}{79}},
  \bibinfo{pages}{031124} (\bibinfo{year}{2009}).

\bibitem[{\citenamefont{Mendoza et~al.}(2013)\citenamefont{Mendoza, Herrmann,
  and Succi}}]{Men13}
\bibinfo{author}{\bibfnamefont{M.}~\bibnamefont{Mendoza}},
  \bibinfo{author}{\bibfnamefont{H.~J.} \bibnamefont{Herrmann}},
  \bibnamefont{and} \bibinfo{author}{\bibfnamefont{S.}~\bibnamefont{Succi}},
  \bibinfo{journal}{Scientific Reports} \textbf{\bibinfo{volume}{3}},
  \bibinfo{pages}{1052} (\bibinfo{year}{2013}).

\bibitem[{\citenamefont{Montakhab et~al.}(2014)\citenamefont{Montakhab,
  Shahsavar, and Ghodrat}}]{Mon14}
\bibinfo{author}{\bibfnamefont{A.}~\bibnamefont{Montakhab}},
  \bibinfo{author}{\bibfnamefont{L.}~\bibnamefont{Shahsavar}},
  \bibnamefont{and} \bibinfo{author}{\bibfnamefont{M.}~\bibnamefont{Ghodrat}},
  \bibinfo{journal}{Physica A} \textbf{\bibinfo{volume}{412}},
  \bibinfo{pages}{32} (\bibinfo{year}{2014}).

\end{thebibliography}

\end{document}